\documentclass[preprintnumbers,showpacs,amsmath,amssymb,floatfix,twocolumn,prd,superscriptaddress,nofootinbib]{revtex4}
\usepackage{graphicx}
\usepackage{epsfig}
\usepackage{bm}
\usepackage{amsfonts}

\begin{document}

\title{Thermodynamics of dark energy interacting with dark matter and radiation}

\author{Mubasher Jamil}
\email{mjamil@camp.nust.edu.pk} \affiliation{Center for Advanced
Mathematics and Physics, National University of Sciences and
Technology,\\ Rawalpindi, 46000, Pakistan}

\author{Emmanuel N. Saridakis }
\email{msaridak@phys.uoa.gr} \affiliation{Department of Physics,
University of Athens, GR-15771 Athens, Greece}

\author{M. R. Setare}
\email{rezakord@ipm.ir} \affiliation{Department of Science, Payame
Noor University, Bijar, Iran }

\begin{abstract}
We investigate the validity of the generalized second law of
thermodynamics, in the cosmological scenario where dark energy
interacts with both dark matter and radiation. Calculating
separately the entropy variation for each fluid component and for
the apparent horizon itself, we show that the generalized second
law is always and generally valid, independently of the specific
interaction form, of the fluids equation-of-state parameters and
of  the background geometry.
 \end{abstract}


 \maketitle

\section{Introduction}

Recent cosmological observations obtained by SNe Ia {\cite{c1}},
WMAP {\cite{c2}}, SDSS {\cite{c3}} and X-ray {\cite{c4}} indicate
that the observable universe experiences an accelerated expansion.
Although the simplest way to explain this behavior is the
consideration of a cosmological constant \cite{c7}, the known
fine-tuning problem \cite{8} led to the dark energy paradigm. The
dynamical nature of dark energy, at least in an effective level,
can originate from a variable cosmological ``constant''
\cite{varcc}, or from various fields, such is a canonical scalar
field (quintessence) \cite{quint}, a phantom field, that is a
scalar field with a negative sign of the kinetic term
\cite{phant}, or the combination of quintessence and phantom in a
unified model named quintom \cite{quintom}. Finally, an interesting attempt to probe the nature of
dark energy according to some basic quantum gravitational
principles is the holographic dark energy paradigm \cite{holoext}
(although the recent developments in Horava gravity could offer a
dark energy candidate with perhaps better quantum gravitational
foundations \cite{Horawa}).

The aforementioned models offer a satisfactory description of the
dark-energy behavior and its observable features. However,
attributing a dynamical nature to dark energy, these scenarios
introduce a new cosmological problem \cite{8}, namely why are the
densities of vacuum energy and dark matter nearly equal today
since these scale very differently during the expansion history.
The elaboration of this ``coincidence'' problem led to the
consideration of generalized versions of the aforementioned models
with the inclusion of a coupling between dark energy and dark
matter. Thus, various forms of ``interacting'' dark energy models
have been constructed in order to fulfil the observational
requirements, including interacting quintessence
\cite{interacting}, interacting phantom \cite{Guo:2004vg},
interacting Chaplygin gas \cite{Jamil:2008fq} and interacting
holographic dark energy
\cite{Wang:2005jx,Kim:2005at,Setare:2007at}. However, one could
alleviate the coincidence problem in the variable cosmological
``constant'' framework, too \cite{varcc.int}. Finally, since the
universe also comprises radiation, one can consider the scenario
in which dark energy interacts with both dark matter and radiation
\cite{Cruz:2008qp}. We mention that the coupling to radiation is
not new, since it is also in the center of ``warm inflation''
paradigm \cite{warminflation}. The motivation behind this extra
interaction is the so-called `triple coincidence problem'
\cite{hamed}, that is it can offer an explanation to the fact that
the radiation energy density is today only three orders of
magnitude smaller than the dark-matter and dark-energy ones,
although it also scales very differently. In particular, given the
dark-energy-dark-matter coupling, we consider that the energy
dissipated in this interaction is transferred to the radiation
component and vice versa.

In the present work we are interested in investigating the
interaction of dark energy with both dark matter and radiation
fluids from the thermodynamic point of view. In particular, we
desire to examine under what conditions the underlying system
obeys the generalized second law of thermodynamics, namely the sum
of entropies of the individual components, including that of the
background, to be positive. The plan of the work is as follows: In
section \ref{Model}, we construct the scenario where dark energy
interacts with dark matter and radiation fluids. In section
\ref{thermo}, considering the universe as a system bounded by the
apparent horizon, we study the generalized second law of
thermodynamics for our cosmological model. Finally, section
\ref{conclusions} is devoted to the summary of the obtained
results.

\section{Dark energy interacting with dark matter and radiation}
\label{Model}

Lets us construct a scenario where dark energy interacts with
both dark-matter and radiation fluids. Throughout the work  we
consider a spatially homogeneous and isotropic universe described
by the Friedmann-Robertson-Walker metric
\begin{equation}
ds^2=-dt^2+a^2(t)\Big( \frac{dr^2}{1-kr^2}+r^2d\Omega^2 \Big),
\end{equation}
where $k$  is $0,1,-1$ for flat, closed and open geometry
respectively. Thus, the first Friedmann equation writes
\begin{equation}
\label{Fri1}
H^2+\frac{k}{a^2}=\frac{1}{3}\left(\rho_{DE}+\rho_M+\rho_\chi\right),
\end{equation}
where $H$ is the Hubble parameter and $\rho_{DE}$, $\rho_{M}$ and
$\rho_{x}$ denote the energy densities for dark energy, dark
matter and radiation respectively. For simplicity, throughout this
work we are using units where $8\pi G=1$. We mention that the
aforementioned framework holds independently of the specific dark
energy description.

In the case where the various constituents of the universe are
allowed to interact, the conservation equations for their energy
densities write:
\begin{eqnarray}
&& \dot\rho_{DE}+3H(1+w_{DE})\rho_{DE}=-Q^\prime
\nonumber\\
&& \dot\rho_M+3H(1+w_M)\rho_M=Q
\nonumber\\
&&\label{2eq1x} \dot\rho_\chi+3H(1+w_\chi)\rho_\chi=Q^\prime-Q,
\end{eqnarray}
where a dot denotes the derivative with respect to cosmic time. In
these equations we have inserted the equation-of-state parameters
of the various cosmological constituents, defined as
$w_i=p_i/\rho_i$, where $p_i$ is the pressure of the corresponding
constituent $i$.

In expressions (\ref{2eq1x}), $Q$ and $Q^\prime$ describe the
interaction terms, which can have an arbitrary form. In addition,
we do not restrict the sign of these terms, that is $Q<0$
corresponds to energy transfer from dark-matter sector to the
other two constituents, $Q^\prime>0$ corresponds to energy
transfer form dark-energy sector to the other two fluids, and
$Q^\prime<Q$ corresponds to energy loss for radiation. Obviously,
in the case $Q=Q^\prime$, we obtain the usual model where dark
energy interacts only with the dark-matter sector
\cite{interacting}. Additionally, note that this case may
effectively appear as a self-conserved dark energy, with a
non-trivial equation of state mimicking quintessence or phantom,
as in the $\Lambda$XCDM scenario \cite{varcc,varcc.int}. Lastly,
concerning the form of interaction one can use many ansatzes
\cite{Chimento:2003iea}, but we prefer to remain as general as
possible.

Finally, it proves convenient to generalize \cite{Kim:2005at} and
construct the equivalent effective uncoupled model, described by:
\begin{eqnarray}
&&\dot\rho_{DE}+3H(1+w_{DE}^{\text{eff}})\rho_{DE}=0\nonumber\\
&&\dot\rho_M+3H(1+w_M^{\text{eff}})\rho_M=0\nonumber\\
&&\dot\rho_\chi+3H(1+w_\chi^{\text{eff}})\rho_\chi=0, \label{eff0}
\end{eqnarray}
where the effective equation-of-state parameters are given by
\begin{eqnarray}
w_{DE}^{\text{eff}}&=&w_{DE}+\frac{Q^\prime}{3H\rho_{DE}}\nonumber\\
w_M^{\text{eff}}&=&w_M-\frac{Q}{3H\rho_M}\nonumber\\
w_\chi^{\text{eff}}&=&w_\chi+\frac{Q-Q^\prime}{3H\rho_\chi}.
\label{eff}
\end{eqnarray}

\section{Generalized Second Law of Thermodynamics} \label{thermo}

In the previous section we presented the cosmological scenario in
which the dark-energy sector interacts with the dark-matter and
radiation ones. In the present section we proceed to an
investigation of its thermodynamic properties, and in particular
of the generalized second thermodynamic law \cite{Unruh:1982ic}.\\

In the literature, when one desires to examine the thermodynamic
behavior of a cosmological scenario, one considers the universe as
a thermodynamical system.  However, it is not trivial what
`volume' must be used, and in particular what `radius', in order
to acquire a consistent description. This subject becomes more
important under the light of use of black-hole physics
\cite{Hawking} in a cosmological framework \cite{Jacobson:1995ab},
that is connecting the `radius' and `area' of the universe with
its temperature and entropy respectively. For a flat geometry,
using the Hubble horizon, which in this case coincides with the
apparent horizon, as the aforementioned universe `radius', one can
extract the Friedmann equations by applying the first law of
thermodynamics \cite{Frolov:2002va}. However, in order to achieve
the same equivalence in a non-flat FRW geometry, one has to use
the apparent horizon, since the use of Hubble horizon (in this
case the two horizons do not coincide) cannot lead to a physical
result \cite{Cai:2005ra}. The dynamical apparent horizon, a
marginally trapped surface with vanishing expansion, is in general
determined by the relation $h^{ij}\partial_i\tilde r\partial _j
\tilde r=0 $, which implies that the vector $\nabla \tilde r$ is
null (or degenerate) on the apparent horizon surface
\cite{Bak:1999hd}. In a metric of the form $
 ds^2=h_{ij}dx^idx^j+\tilde r^2d\Omega_2^2$, with $h_{ij}=\text{diag}(-1,a^2/(1-kr^2))$, $i,j=0,1
$, it writes \cite{Bak:1999hd}:
\begin{equation}
\label{apphor}
 \tilde{r}_A=\frac{1}{\sqrt{H^2+\frac{k}{a^2}}}.
\end{equation}
Furthermore, for a dynamical spacetime the apparent horizon has
been argued to be a causal horizon and it is associated with the
gravitational entropy and surface gravity
\cite{Hayward:1997jp,Bak:1999hd}. Therefore, for the purpose of
this work, we consider the universe as a thermodynamical system
with the apparent horizon surface being its boundary.\\

The main goal of the present work is to examine the validity of
the generalized second law of thermodynamics. As we have already
mentioned, the Friedmann equations themselves arise straightaway
from the first law of thermodynamics. Although this has been shown
in the literature for only one fluid
\cite{Hayward:1997jp,Bak:1999hd,Cai:2005ra,Sheykhi:2007zp}, it can
be easily extended to the case at hand. Thus, one could
equivalently construct the scenario of section \ref{Model} based
solely on thermodynamics, with the only external input still
needed, in order to close the equations system, being the
information about the fluid content of the universe, and in
particular the (three in our case) conservation equations
(\ref{2eq1x}).

Let us now proceed to the investigation of the generalized second
law of thermodynamics in the universe. We are going to examine
whether the sum of the entropy enclosed by the apparent horizon
and the entropy of the apparent horizon itself, is not a
decreasing function of time. Simple arguments suggest that after
equilibrium establishes and the universe background geometry
becomes FRW, all the fluids in the universe acquire the same
temperature $T$ \cite{Setare:2007at}, which is moreover equal to
the temperature of the horizon $T_h$
\cite{Frolov:2002va,Cai:2005ra}, otherwise the energy flow would
deform this geometry \cite{pa}.

In general, the apparent horizon $\tilde{r}_A$ is a function of
time. Thus, a change $d\tilde {r}_A$ in time $dt$ will lead to a
volume-change $dV$, while the energy and entropy will change by
$dE$ and $dS$ respectively. However, since in the two states there
is a common source $T_{\mu\nu}$, we can consider that the pressure
$P$ and the temperature $T$ remain the same
\cite{Frolov:2002va,Cai:2005ra}. Such a consideration is standard
in thermodynamics, where one considers two equilibrium states
differing infinitesimally in the extensive variables like entropy,
energy and volume, while having the same values for the intensive
variables like temperature and pressure.
 In this case the first law of
thermodynamics writes $TdS=dE+PdV$, and therefore the  dark-energy
and  dark-matter entropies read \cite{abdalla}:
\begin{eqnarray}
\label{en1}
dS_{DE}&=&\frac{1}{T}\Big(P_{DE} dV+dE_{DE}\Big)\nonumber\\
\label{en2}dS_M&=&\frac{1}{T}\Big(P_M dV+dE_M\Big)\nonumber\\
\label{en3}dS_\chi&=&\frac{1}{T}\Big(P_\chi dV+dE_\chi\Big),
\end{eqnarray}
 where $V=4 \pi \tilde{r}_A^3/3$ is the volume of the system bounded by the
 apparent horizon and thus $ dV=4\pi\tilde{r}_A^2d\tilde
{r}_A$. In the aforementioned expressions we have also added the
 corresponding relation for radiation. We mention that due to
 equilibration, all the constituents (fluids)
 of the universe have the same temperature, while their energy and
 pressure are in general different.  Finally, it proves useful to
 divide (\ref{en3}) by $dt$, obtaining:
\begin{eqnarray}
\dot{S}_{DE}&=&\frac{1}{T}\Big(P_{DE}\,
4\pi\tilde{r}_A^2\dot{\tilde
{r}}_A+\dot{E}_{DE}\Big)\nonumber\\
\dot{S}_M&=&\frac{1}{T}\Big(P_M \, 4\pi\tilde{r}_A^2\dot{\tilde
{r}}_A+\dot{E}_M\Big)\nonumber\\
\label{en3bb}\dot{S}_\chi&=&\frac{1}{T}\Big(P_\chi \,
4\pi\tilde{r}_A^2\dot{\tilde {r}}_A+\dot{E}_\chi\Big),
\end{eqnarray}
where
\begin{equation}
\dot{\tilde
r}_A=\frac{1}{2}H\tilde{r}_A^3\Big[(1+w_{DE})\rho_{DE}+(1+w_M)\rho_M+(1+w_\chi)\rho_\chi\Big],
\label{dotrh}
\end{equation}
as it easily arises differentiating the Friedmann equation $
\frac{1}{\tilde{r}_A^2}=\frac{1}{3}(\rho_{DE}+\rho_M+\rho_\chi)$
and using  (\ref{2eq1x}).

In order to connect the thermodynamically relevant quantities,
namely the energies $E_i$ and pressures $P_i$, with the
cosmologically relevant ones, namely the energy densities $\rho_i$
and the pressures $p_i$, we can straightforwardly use:
\begin{eqnarray}
E_{DE}&=&\frac{4\pi}{3}\tilde{r}_A^3\rho_{DE}\nonumber\\
E_M&=&\frac{4\pi}{3}\tilde{r}_A^3\rho_M \nonumber\\
E_\chi&=&\frac{4\pi}{3}\tilde{r}_A^3\rho_\chi, \label{Eitilde}
\end{eqnarray}
and
\begin{eqnarray}
P_{DE}&=&w_{DE}^{\text{eff}}\rho_{DE}\nonumber\\
 P_M&=&w_M^{\text{eff}}\rho_M\nonumber\\
  P_\chi&=&w_\chi^{\text{eff}}\rho_\chi.
  \label{Pilde}
\end{eqnarray}
Note that expressions (\ref{Pilde}) arise necessarily from the
``uncoupled'' form of the system (relations (\ref{eff0}) and
(\ref{eff})), since one needs an (effective or not)
non-interacting system in order to apply basic thermodynamics and
avoid concepts like the chemical potential. Inserting the
time-derivatives of (\ref{Eitilde}), along with (\ref{Pilde}),
into (\ref{en3bb}), and using (\ref{2eq1x}), we obtain:
\begin{eqnarray}
\dot{S}_{DE}&=&\frac{1}{T}\left(1+w_{DE}^{\text{eff}}\right)\rho_{DE}\,
4\pi\tilde{r}_A^2\left(\dot{\tilde
{r}}_A-H\tilde{r}_A\right)\label{sdotfluids1}\\
\dot{S}_M&=&\frac{1}{T}\left(1+w_{M}^{\text{eff}}\right)\rho_{M}\,
4\pi\tilde{r}_A^2\left(\dot{\tilde
{r}}_A-H\tilde{r}_A\right)\label{sdotfluids2}\\
\dot{S}_\chi&=&\frac{1}{T}\left(1+w_{\chi}^{\text{eff}}\right)\rho_\chi\,
4\pi\tilde{r}_A^2\left(\dot{\tilde
{r}}_A-H\tilde{r}_A\right).\label{sdotfluids3}
\end{eqnarray}

At this stage, we have to connect the temperature of the fluids
$T$, which is equal to that of the horizon $T_h$, with the
geometry of the universe. According to the generalization of black
hole thermodynamics \cite{Hawking} to a cosmological framework,
the temperature of the horizon is related to its radius through
\cite{Jacobson:1995ab,Cai:2005ra}
\begin{equation}
\label{Th}
 T_h=\frac{1}{2\pi\tilde{r}_A}.
\end{equation}
Finally, concerning the entropy of the horizon, one can define it
as \cite{Jacobson:1995ab,Cai:2005ra} $S_h=4\pi\tilde{r}_A^2/(4G)$,
and since in this work we are using units where $8\pi G=1$, we
acquire $S_h=8\pi^2\tilde{r}_A^2$. Hence, we obtain:
\begin{equation}
\label{Sdoth}
 \dot{S}_{h}=16\pi^2\tilde{r}_A\dot{\tilde {r}}_A.
\end{equation}

Let us now proceed to the calculation of the total entropy
variation. Adding relations
(\ref{sdotfluids1})-(\ref{sdotfluids3}) and (\ref{Sdoth}), and
substituting  the effective equation-of-state parameters through
(\ref{eff}), we find:
\begin{widetext}
\begin{eqnarray}
\label{Sdottot}
 \dot{S}_{tot}\equiv\dot{S}_{DE}+\dot{S}_M+\dot{S}_\chi+ \dot{S}_h=
8\pi^2\tilde{r}_A^3\left(\dot{\tilde
{r}}_A-H\tilde{r}_A\right)\Big[(1+w_{DE})\rho_{DE}+(1+w_M)\rho_M+(1+w_\chi)\rho_\chi\Big]+16\pi^2\tilde{r}_A\dot{\tilde
{r}}_A.
\end{eqnarray}
\end{widetext}
Note that the aforementioned expression was simplified due to the
useful  relation
\begin{eqnarray}(1+w_{DE}^{\text{eff}})\rho_{DE}+(1+w_M^{\text{eff}})\rho_M+(1+w_\chi^{\text{eff}})\rho_\chi=\nonumber\\
(1+w_{DE})\rho_{DE}+(1+w_M)\rho_M+(1+w_\chi)\rho_\chi.
\end{eqnarray}
Thus, substituting also $\dot{\tilde {r}}_A$ by (\ref{dotrh}) we
result to:
\begin{widetext}
\begin{eqnarray}
\label{Sdottot2}
 \dot{S}_{tot}=
4\pi^2\tilde{r}_A^6H\Big[(1+w_{DE})\rho_{DE}+(1+w_M)\rho_M+(1+w_\chi)\rho_\chi\Big]^2\geq0.
\end{eqnarray}
\end{widetext}
 (\ref{Sdottot2}) provides the expression of the generalized
second law of thermodynamics in the scenario where dark energy
interacts with dark matter and radiation. The fact that $
\dot{S}_{tot}$ is always non-negative proves the validity of this
law at all cosmological times. We mention that this result holds
independently of the interaction form, of the fluids
equation-of-state parameters, and of the background geometry,
provided it is FRW. Finally, note that in the aforementioned
analysis we have not taken into account the possible black-hole
formation in the universe and its effect on entropy.

Let us make a  comment here for completeness. As we observe from
(\ref{Sdottot2}), there is a critical value of the dark-energy
equation-of-state parameter that leads to $ \dot{S}_{tot}=0$. In
particular this happens if
$w_{DE{cr}}=-1-(1+w_M)\rho_M/\rho_{DE}-(1+w_\chi)\rho_\chi/\rho_{DE}$.
That is, for regular dark-matter and radiation, $w_{DE{cr}}$ lies
in the phantom regime. However, in order for $w_{DE}$ to obtain
this value one definitely needs a short of fine tuning. Its
calculation at present (where the corresponding density parameters
are $\Omega_M\approx0.72$, $\Omega_M\approx0.28$, $w_M\approx0$
and neglecting radiation for simplicity) leads to
$w_{DEcr}\approx-1.39$, which is outside the observational
intervals for $w_{DE}$. But this could change in the future, if
$\Omega_M$ decrease sufficiently. We mention that if $w_{DE}$ is
even smaller, then $ \dot{S}_{tot}$ will become positive again.
Finally, from (\ref{dotrh}) we observe that this value for
$w_{DE}=w_{DEcr}$ leads also to $\dot{\tilde r}_A=0$. Thus, from
the apparent horizon definition (\ref{apphor}) we can easily
calculate the corresponding solution for the scale factor, namely
$
 a(t)=e^{\alpha t+\beta}+4k e^{-\alpha t-\beta}$ and $
  a(t)=e^{-\alpha t-\beta}+4k e^{\alpha t+\beta}$,
where $\alpha,\beta$ are constants, and with the second solution
existing only for a closed universe. Obviously, for large times
these solution tend to a de Sitter solution. Note however that if
there is a strong energy-transfer from dark energy to the other
sectors, the former will not necessarily dominate the universe
completely.

In the following we discuss about the sign of the temperature and
entropy. In general, if the total equation-of-state parameter of
the universe lies above the phantom divide, then both these
quantities are unambiguously positive as usual. However, if the
universe lies in the phantom phase the subject is still open in
the literature. Assuming a zero chemical potential the temperature
must be negative, with the density and the entropy positive
\cite{phantomtherm} and assuming a negative chemical potential
then temperature, entropy and density are positive
\cite{Lima:2008uk}. Additionally, in \cite{Saridakis:2009uu} it
was shown that even with an arbitrary chemical potential the
temperature of a phantom universe is negative, with the density
and the entropy positive, while in \cite{pa} it was found that the
phantom temperature is positive and its entropy negative. Finally,
in  \cite{Nojiri:2004pf} it was argued that one can describe the
phantom universe either with negative temperature and positive
entropy, or with negative entropy and positive temperature. In the
analysis of this work equilibrium requires all the fluids in the
universe to have the same temperature, which is moreover equal to
the temperature of the horizon (although one could put into
question the equilibrium assumption and examine the
non-equilibrium case too). Since the horizon temperature is always
positive, it is deduced that the universe temperature will be
positive even if it lies in the phantom phase. Thus, in order to
be in agreement with the literature, one should have a negative
universe entropy in this case. Note however that the negative
entropy of the universe ingredients is overcome by the positive
horizon entropy, and thus the total entropy is always positive.

 In
particular,  and assuming complete dark energy domination for
simplicity, in the quintessence regime we have $\dot{\tilde
r}_A>0$ (according to (\ref{dotrh})). Thus, according to
(\ref{Sdoth}) we obtain $\dot{S}_h>0$, while according to
(\ref{sdotfluids1}) we have $\dot{S}_{DE}<0$. Furthermore, $S_h>0$
(as always), while $S_{DE}>0$ (since we are in the quintessence
regime). Thus, in this regime we have $\dot{S}_{tot}>0$ (according
to (\ref{Sdottot2})) and $S_{tot}>0$.  On the other hand, in the
phantom regime (\ref{dotrh}) leads to $\dot{\tilde r}_A<0$, and
thus (\ref{Sdoth}) gives $\dot{S}_h<0$, while (\ref{sdotfluids1})
gives $\dot{S}_{DE}>0$. Moreover, $S_h>0$ (as always), while
$S_{DE}<0$ (since we are in the phantom regime). However, in this
case we obtain $S_{tot}>0$ and also $\dot{S}_{tot}>0$. So in
summary, the present work and the validity of the generalized
second law, is consistent with the positive-temperature and
negative-entropy picture of the phantom dark energy. Finally, note
that the inconsistency of the generalized second law with the
negative-temperature and positive-entropy picture of the phantom
dark energy, was already mentioned in
\cite{MohseniSadjadi:2005ps}.

Before closing this section let us make a comment on the horizon
we use in the present work. Concerning the first law of
thermodynamics it has been proven in \cite{Padmanabhan:2009ry}
that the  field equations of any (diffeomorphism invariant)
gravitational theory
 must be expressible as a thermodynamic identity, $TdS =
dE$, around any event in the spacetime. However, concerning the
generalized second law of thermodynamics, it has been shown that
it is generally valid only if one uses the apparent horizon
\cite{Hayward:1997jp,Bak:1999hd,Cai:2005ra,Sheykhi:2007zp}, while
it is conditionally valid for other horizon choices
\cite{MohseniSadjadi:2005ps}. Thus, repeating our calculations
using the future event horizon $R_h=\int^\infty_a da/(Ha^2)$
instead of the apparent horizon $r_A$, we find that the
generalized second law is not always valid.
 Clearly, the separate investigation of each horizon is a crucial
 and open subject in the thermodynamic aspects of gravity, as it
 is discussed in detail in the recent review
 \cite{Padmanabhan:2009ry.bb}.

\section{Conclusions}\label{conclusions}

In this work we investigated the cosmological scenario where dark
energy interacts with both dark matter and radiation, a scenario
which could alleviate the triple coincidence problem \cite{hamed}.
After reminding that in such fluid cosmological models the
Friedmann equations can arise from the first law of
thermodynamics, with the fluid conservation equations being the
only external input required to close the equations system, we
examined the validity of the generalized second law of
thermodynamics. Considering the universe as a thermodynamical
system bounded by the apparent horizon, and calculating separately
the entropy variation for each fluid component and for the horizon
itself, we resulted to an expression for the time derivative of
the total entropy of the universe.

According to our main result, that is expression (\ref{Sdottot2}),
the time derivative of the total entropy is always non-negative,
and this holds independently of the specific interaction form, of
the fluids equation-of-state parameters, and of the background
geometry. Thus, the generalized second law of thermodynamics is
always and generally valid, as long as one considers the apparent
horizon as the universe ``radius'' (the use of other choices, such
is the future event horizon, leads to conditional validity only).
However, we mention that
 the present work  is consistent with the
positive-temperature and negative-entropy picture of the phantom
dark energy \cite{Nojiri:2004pf}. Finally, it is interesting to
notice that if the dark-energy equation-of-state parameter takes a
critical phantom value, then the total entropy of the universe
remains constant.

In the present work we have proven the validity of the generalized
second law of thermodynamic in the scenario where dark energy
interacts with dark-matter  and radiation sectors. This result is
necessary for the consideration of such scenarios, but it is not
sufficient. One must also examine whether such models affect the
known cosmological epochs, before proceeding to their safe use.
But such a study is beyond the purpose of this work and it is left
for future investigation.

\begin{acknowledgments}
The authors would like to thank R. Horvat and H. Mohseni Sadjadi
for useful discussions, and an anonymous referee for fruitful
comments and advices.
\end{acknowledgments}


\begin{thebibliography}{99}

\bibitem{c1}
 A.~G.~Riess {\it et al.}  [Supernova Search Team Collaboration],
  Astron.\ J.\  {\bf 116}, 1009 (1998);
S. Perlmutter {\it et al.} [Supernova Cosmology Project
Collaboration], Astrophys. J. {\bf 517}, 565 (1999).

\bibitem{c2}
C. L. Bennett {\it et al.}, Astrophys. J. Suppl. {\bf 148}, 1
(2003).
\bibitem{c3}
M. Tegmark {\it et al.} [SDSS Collaboration], Phys. Rev. D {\bf
69}, 103501 (2004).
\bibitem{c4}
S. W. Allen, \emph{et al.}, Mon. Not. Roy. Astron. Soc. {\bf 353},
457 (2004).

\bibitem{c7}
V. Sahni and A. Starobinsky, Int. J. Mod. Phy. D {\bf 9}, 373
(2000); P. J. Peebles and B. Ratra, Rev. Mod. Phys. {\bf 75}, 559
(2003).

\bibitem{8}P.~J.~Steinhardt,  {\it {Critical Problems in Physics}} (1997), Princeton
University Press.


\bibitem{varcc}
J.~Sola and H.~Stefancic,
Phys.\ Lett.\  B {\bf 624}, 147 (2005);
J.~Sola and H.~Stefancic,
Mod.\ Phys.\ Lett.\  A {\bf 21}, 479 (2006);
I.~L.~Shapiro and J.~Sola,
Phys.\ Lett.\  B {\bf 682}, 105 (2009).


\bibitem{quint}
B.~Ratra and P.~J.~E.~Peebles, Phys.\ Rev.\ D {\bf 37}, 3406
(1988); C.~Wetterich, Nucl.\ Phys.\ B {\bf 302}, 668 (1988);
A.~R.~Liddle and R.~J.~Scherrer, Phys.\ Rev.\ D {\bf 59}, 023509
(1999); I.~Zlatev, L.~M.~Wang and P.~J.~Steinhardt, Phys.\ Rev.\
Lett.\ {\bf 82}, 896 (1999); Z.~K.~Guo, N.~Ohta and Y.~Z.~Zhang,
Mod.\ Phys.\ Lett.\  A {\bf 22}, 883 (2007);
  S.~Dutta, E.~N.~Saridakis and R.~J.~Scherrer,
Phys.\ Rev.\  D {\bf 79}, 103005 (2009).

\bibitem{phant} R. R. Caldwell, Phys.
Lett. B {\bf{545}}, 23 (2002); R.~R.~Caldwell, M.~Kamionkowski and
N.~N.~Weinberg, Phys. Rev. Lett. {\bf 91}, 071301 (2003); S.
Nojiri and S. D. Odintsov, Phys. Lett. B {\bf 562}, 147 (2003); V.
K. Onemli and R. P. Woodard, Phys.\ Rev.\ D {\bf 70}, 107301
(2004); M. R. Setare, J. Sadeghi, A. R. Amani, Phys. Lett. B {\bf
666}, 288, (2008);
  M.~R.~Setare and E.~N.~Saridakis,
  JCAP {\bf 0903}, 002 (2009);
  E.~N.~Saridakis,
  Nucl.\ Phys.\  B {\bf 819}, 116 (2009).

\bibitem{quintom}
B.~Feng, X.~L.~Wang and X.~M.~Zhang, Phys.\ Lett.\  B {\bf 607},
35 (2005);
Z. K. Guo, {\it{et al.}}, Phys. Lett. B {\bf 608}, 177 (2005);
M.-Z Li, B. Feng, X.-M Zhang, JCAP, 0512, 002 (2005); B. Feng, M.
Li, Y.-S. Piao and X. Zhang, Phys. Lett. B {\bf 634}, 101 (2006);
M. R. Setare, Phys. Lett. B {\bf 641}, 130 (2006); W. Zhao and Y.
Zhang, Phys. Rev. D {\bf73}, 123509 (2006);
 M. R.
Setare, J. Sadeghi, and A. R. Amani, Phys. Lett. B {\bf 660}, 299
(2008);
  M.~R.~Setare and E.~N.~Saridakis,
  Phys.\ Lett.\  B {\bf 668}, 177 (2008);
  M.~R.~Setare and E.~N.~Saridakis,
  JCAP {\bf 0809}, 026 (2008);
  M.~R.~Setare and E.~N.~Saridakis,
  Int.\ J.\ Mod.\ Phys.\  D {\bf 18}, 549 (2009).

\bibitem{holoext}
  S.~D.~H.~Hsu,
  Phys.\ Lett.\ B {\bf 594}, 13 (2004);
  M.~Li,
  Phys.\ Lett.\ B {\bf 603}, 1 (2004);
   Q.~G.~Huang and M.~Li,
  JCAP {\bf 0408}, 013 (2004);
 M.~Ito,
 Europhys.\ Lett.\  {\bf 71}, 712 (2005);
    X.~Zhang and F.~Q.~Wu,
  Phys.\ Rev.\  D {\bf 72}, 043524 (2005);
  D.~Pavon and W.~Zimdahl,
  Phys.\ Lett.\ B {\bf 628}, 206 (2005);
  S.~Nojiri and S.~D.~Odintsov,
  Gen.\ Rel.\ Grav.\  {\bf 38}, 1285 (2006);
  E.~Elizalde, S.~Nojiri, S.~D.~Odintsov and P.~Wang,
  Phys.\ Rev.\ D {\bf 71}, 103504 (2005);
  H.~Li, Z.~K.~Guo and Y.~Z.~Zhang,
  Int.\ J.\ Mod.\ Phys.\ D {\bf 15}, 869 (2006);
  E.~N.~Saridakis,
  Phys.\ Lett.\  B {\bf 660}, 138 (2008);
  E.~N.~Saridakis,
  JCAP {\bf 0804}, 020 (2008);
  E.~N.~Saridakis,
  Phys.\ Lett.\  B {\bf 661}, 335 (2008).







\bibitem{Horawa}
  P.~Horava,
  Phys.\ Rev.\  D {\bf 79}, 084008 (2009);
  G.~Calcagni,
  arXiv:0904.0829 [hep-th];
    E.~Kiritsis and G.~Kofinas,
  Nucl.\ Phys.\  B {\bf 821}, 467 (2009);
  H.~Lu, J.~Mei and C.~N.~Pope,
  arXiv:0904.1595 [hep-th];
   C.~Charmousis, G.~Niz, A.~Padilla and P.~M.~Saffin,
  arXiv:0905.2579 [hep-th];
  E.~N.~Saridakis,
  arXiv:0905.3532 [hep-th];
  X.~Gao, Y.~Wang, R.~Brandenberger and A.~Riotto,
  arXiv:0905.3821 [hep-th];
  M.~i.~Park,
  arXiv:0905.4480 [hep-th];
  Y.~F.~Cai and E.~N.~Saridakis,
  arXiv:0906.1789 [hep-th];
  M.~Botta-Cantcheff, N.~Grandi and M.~Sturla,
  arXiv:0906.0582 [hep-th];
  M.~R.~Setare,
  arXiv:0909.0456 [hep-th];
  C.~Germani, A.~Kehagias and K.~Sfetsos,
  JHEP {\bf 0909}, 060 (2009);
  G.~Leon and E.~N.~Saridakis,
  JCAP {\bf 0911}, 006 (2009).



\bibitem{interacting}
  C.~Wetterich,
  Astron.\ Astrophys.\  {\bf 301}, 321 (1995)
  L.~Amendola,
  Phys.\ Rev.\  D {\bf 60}, 043501 (1999);
  A.~P.~Billyard and A.~A.~Coley,
  Phys.\ Rev.\  D {\bf 61}, 083503 (2000);
    A.~Nunes, J.~P.~Mimoso and T.~C.~Charters,
  Phys.\ Rev.\  D {\bf 63}, 083506 (2001);
  G.~R.~Farrar and P.~J.~E.~Peebles,
  Astrophys.\ J.\  {\bf 604}, 1 (2004);
    D.~F.~Mota and C.~van de Bruck,
  Astron.\ Astrophys.\  {\bf 421}, 71 (2004);
   X.~Zhang,
  Mod.\ Phys.\ Lett.\  A {\bf 20}, 2575 (2005);
 T.~Gonzalez, G.~Leon and I.~Quiros,
  Class.\ Quant.\ Grav.\  {\bf 23}, 3165 (2006);
    T.~Clifton and J.~D.~Barrow,
  Phys.\ Rev.\  D {\bf 73}, 104022 (2006);
  M.~Manera and D.~F.~Mota,
  Mon.\ Not.\ Roy.\ Astron.\ Soc.\  {\bf 371}, 1373 (2006);
    T.~Clifton and J.~D.~Barrow,
  Phys.\ Rev.\  D {\bf 75}, 043515 (2007).








\bibitem{Guo:2004vg}
  Z.~K.~Guo, R.~G.~Cai and Y.~Z.~Zhang,
  JCAP {\bf 0505}, 002 (2005);
  Z.~K.~Guo and Y.~Z.~Zhang,
  Phys.\ Rev.\  D {\bf 71}, 023501 (2005);
R.~Curbelo, T.~Gonzalez, G. Leon and I.~Quiros,
  Class.\ Quant.\ Grav.\  {\bf 23}, 1585 (2006);
          J.~D.~Barrow and T.~Clifton,
  Phys.\ Rev.\  D {\bf 73}, 103520 (2006)
  T.~Gonzalez and I.~Quiros,
  Class.\ Quant.\ Grav.\  {\bf 25}, 175019 (2008);
  X.~m.~Chen, Y.~g.~Gong and E.~N.~Saridakis,
  JCAP {\bf 0904}, 001 (2009).

\bibitem{Jamil:2008fq}
  H.~Garcia-Compean, G.~Garcia-Jimenez, O.~Obregon and C.~Ramirez,
  JCAP {\bf 0807}, 016 (2008);
  M.~Jamil and M.~A.~Rashid,
  Eur.\ Phys.\ J.\  C {\bf 60}, 141 (2009).

\bibitem{Wang:2005jx}
  B.~Wang, Y.~G.~Gong and E.~Abdalla,
  Phys.\ Lett.\  B {\bf 624}, 141 (2005);
    M.~R.~Setare,
  Phys.\ Lett.\  B {\bf 642}, 1 (2006);
  B.~Hu and Y.~Ling,
  Phys.\ Rev.\ D {\bf 73}, 123510 (2006).

\bibitem{Kim:2005at}
H.~Kim, H.~W.~Lee and Y.~S.~Myung,
  Phys.\ Lett.\ B {\bf 632}, 605 (2006).

\bibitem{Setare:2007at}
  M.~R.~Setare,
  JCAP {\bf 0701}, 023 (2007).

\bibitem{varcc.int}
  J.~Grande, J.~Sola and H.~Stefancic,
  JCAP {\bf 0608}, 011 (2006);
  J.~Grande, A.~Pelinson and J.~Sola,
  Phys.\ Rev.\  D {\bf 79}, 043006 (2009).


\bibitem{Cruz:2008qp}
  N.~Cruz, S.~Lepe and F.~Pena,
  Phys.\ Lett.\  B {\bf 663}, 338 (2008).

\bibitem{warminflation}
  A.~Berera,
  Phys.\ Rev.\ Lett.\  {\bf 75}, 3218 (1995);
  A.~Berera, M.~Gleiser and R.~O.~Ramos,
  Phys.\ Rev.\ Lett.\  {\bf 83}, 264 (1999);
  J.~M.~F.~Maia and J.~A.~S.~Lima,
  Phys.\ Rev.\  D {\bf 60}, 101301 (1999);
  J.~P.~Mimoso, A.~Nunes and D.~Pavon,
  Phys.\ Rev.\  D {\bf 73}, 023502 (2006).

\bibitem{hamed}
  N.~Arkani-Hamed, L.~J.~Hall, C.~F.~Kolda and H.~Murayama,
  Phys.\ Rev.\ Lett.\  {\bf 85}, 4434 (2000).

\bibitem{Chimento:2003iea}
  L.~P.~Chimento, A.~S.~Jakubi, D.~Pavon and W.~Zimdahl,
  Phys.\ Rev.\  D {\bf 67}, 083513 (2003);
  C.~G.~B\"{o}hmer, G.~Caldera-Cabral, R.~Lazkoz and R.~Maartens,
  Phys.\ Rev.\  D {\bf 78}, 023505 (2008).

\bibitem{Unruh:1982ic}
  W.~G.~Unruh and R.~M.~Wald,
  Phys.\ Rev.\  D {\bf 25}, 942 (1982);
  P.~C.~W.~Davies,
  Class.\ Quant.\ Grav.\  {\bf 4}, L225 (1987);
    H.~Mohseni Sadjadi,
  Phys.\ Rev.\  D {\bf 76}, 104024 (2007);
    R.~Horvat,
  Phys.\ Lett.\  B {\bf 648}, 374 (2007);
  M.~R.~Setare and E.~C.~Vagenas,
  Phys.\ Lett.\  B {\bf 666}, 111 (2008);
  A.~Sheykhi and B.~Wang,
  Phys.\ Lett.\  B {\bf 678}, 434 (2009).






\bibitem{Hawking}
G. W. Gibbons and S. W. Hawking, Phys. Rev. D {\bf15}, 2738
(1977).

\bibitem{Jacobson:1995ab}
  T.~Jacobson,
  Phys.\ Rev.\ Lett.\  {\bf 75}, 1260 (1995);
  T.~Padmanabhan,
  Phys.\ Rept.\  {\bf 406}, 49 (2005);
  A.~Paranjape, S.~Sarkar and T.~Padmanabhan,
  Phys.\ Rev.\  D {\bf 74}, 104015 (2006).




\bibitem{Frolov:2002va}
  A.~V.~Frolov and L.~Kofman,
  JCAP {\bf 0305}, 009 (2003);
  U.~H.~Danielsson,
  Phys.\ Rev.\  D {\bf 71}, 023516 (2005);
  R.~Bousso,
  Phys.\ Rev.\  D {\bf 71}, 064024 (2005).

\bibitem{Cai:2005ra}
  R.~G.~Cai and S.~P.~Kim,
  JHEP {\bf 0502}, 050 (2005);
M.~Akbar and R.~G.~Cai,
  Phys.\ Rev.\  D {\bf 75}, 084003 (2007).

\bibitem{Bak:1999hd}
  D.~Bak and S.~J.~Rey,
  Class.\ Quant.\ Grav.\  {\bf 17}, L83 (2000).

\bibitem{Hayward:1997jp}
  S.~A.~Hayward,
  Class.\ Quant.\ Grav.\  {\bf 15}, 3147 (1998);
  S.~A.~Hayward, S.~Mukohyama and M.~C.~Ashworth,
  Phys.\ Lett.\  A {\bf 256}, 347 (1999).

\bibitem{Sheykhi:2007zp}
  A.~Sheykhi, B.~Wang and R.~G.~Cai,
  Nucl.\ Phys.\  B {\bf 779}, 1 (2007).

\bibitem{pa}
  G.~Izquierdo and D.~Pavon,
  Phys.\ Lett.\  B {\bf 633}, 420 (2006).

\bibitem{abdalla}
  B.~Wang, Y.~Gong and E.~Abdalla,
  Phys.\ Rev.\  D {\bf 74}, 083520 (2006);
  Y.~Gong, B.~Wang and A.~Wang,
  Phys.\ Rev.\  D {\bf 75}, 123516 (2007).


\bibitem{phantomtherm}
  P.~F.~Gonzalez-Diaz and C.~L.~Siguenza,
  Nucl.\ Phys.\  B {\bf 697}, 363 (2004);
  Y.~S.~Myung,
  Phys.\ Lett.\  B {\bf 671}, 216 (2009).


\bibitem{Lima:2008uk}
  J.~A.~S.~Lima and S.~H.~Pereira,
  Phys.\ Rev.\  D {\bf 78}, 083504 (2008);
  S.~H.~Pereira and J.~A.~S.~Lima,
  Phys.\ Lett.\  B {\bf 669}, 266 (2008).

\bibitem{Saridakis:2009uu}
  E.~N.~Saridakis, P.~F.~Gonzalez-Diaz and C.~L.~Siguenza,
  Class.\ Quant.\ Grav.\  {\bf 26}, 165003 (2009).

\bibitem{Nojiri:2004pf}
  I.~H.~Brevik, S.~Nojiri, S.~D.~Odintsov and L.~Vanzo,
  Phys.\ Rev.\  D {\bf 70}, 043520 (2004);
  S.~Nojiri and S.~D.~Odintsov,
  Phys.\ Rev.\  D {\bf 70}, 103522 (2004).

\bibitem{MohseniSadjadi:2005ps}
  H.~Mohseni Sadjadi,
  Phys.\ Rev.\  D {\bf 73}, 063525 (2006);
  R.~Horvat,
  Phys.\ Lett.\  B {\bf 664}, 201 (2008).

\bibitem{Padmanabhan:2009ry}
  T.~Padmanabhan,
  arXiv:0903.1254 [hep-th].

\bibitem{Padmanabhan:2009ry.bb}
  T.~Padmanabhan,
  arXiv:0911.5004 [gr-qc].


\end{thebibliography}
\end{document}